\documentclass[letter]{jpsj3} 

\usepackage{mathrsfs} %
\usepackage{amsmath,amssymb}
\def\mps#1{\mathbf{#1}}
\def\PsiA{A}

\def\i{i}
\def\sop{{\bf s}}
\def\sy{{\bf s}^y}

\def\ua{\uparrow}
\def\da{\downarrow}

\newcommand{\ket}[1]{\left| #1 \right\rangle}

\newcommand{\bra}[1]{\left\langle #1 \right|}

\def\ua{\uparrow}
\def\da{\downarrow}

\def\Tr{{\rm Tr\,}}
\def\bege{\begin{equation}}
\def\ende{\end{equation}}

\def\e#1{\,{\rm e}#1}

\def\vec#1{{\overrightarrow{#1}}}
\usepackage[]{bm}
\def\vec#1{{\bm{#1}}}

\def\Tr{\mathop{{\rm Tr}}}

\def\and{\,\,\&\,}

\def\ket#1{|#1\rangle }
\def\bra#1{\langle #1|}
\def\braket#1#2{\langle #1|#2\rangle}

\def\i{{\bf i}}

\def\skipm#1{}
\def\refeq#1{Eq.(\ref{#1})}

\def\vec#1{{\mathbf{#1}}}
\def\section#1{\par}
\def\i{i}
\def\mps#1{\mathbf{#1}}
\def\PsiA{\mps{\Psi}}
\def\UKT{\hat{U}}

\def\tot{} 
\def\rml{{\ell}} 
\def\singlet{{\rm s}} 
\title{Natural extension of hidden $Z_2 \times Z_2$ symmetry toward arbitrary integer spin chains
}
\author{Isao \textsc{Maruyama}\thanks{E-mail address: maru@mp.es.osaka-u.ac.jp}}

\inst{Graduate School of Engineering Science, Osaka University, 
Toyonaka, Osaka 560-8531, Japan \\
}
\abst{We show how entangled valence-bond singlet pairs are disentangled partially and totally 
by the Kennedy-Tasaki transformation which reveals the hidden $Z_2\times Z_2$ symmetry
in valence-bond-solid chains
as a higher-spin generalization of the previous studies 
toward the intermediate-$D$ state.
The totally disentangled states correspond to
four Ising-like states with $Z_2$ variables on the boundary.
We present a simple expression of results by using
 the spin decomposition and the boundary matrix.
}
\kword{hidden $Z_2 \times Z_2$ symmetry, valence-bond solid, matrix product state, disentangler}

\begin{document}
\maketitle

\section{Introduction}
Haldane\cite{PLA.93.464} predicted that
the ground state in integer spin quantum antiferromagnetic chains
is unique, massive, and disordered.
This conjecture has been examined by many numerical, experimental,
and rigorous studies.
Here, ``massive'' means a gapped ground state with the Haldane gap, and
``disordered'' indicates exponentially decaying correlation functions.
In spite of disorder,
there is
the hidden antiferromagnetic order
detected by the nonlocal string order parameter\cite{PRB.40.4709}.
It is regarded as consequence of the hidden $Z_2 \times Z_2$ symmetry breaking\cite{PRB.45.304}.

This hidden order of the disordered ground state is understood by the restricted solid-on-solid(RSOS) model\cite{PRB.40.4709}.
To illustrate it,
let us consider a spin-1/2 Ising (or $Z_2$) variable $\sigma$, where $\sigma=\ua,$ or $\da$.
When we randomly generate $\sigma_i$ on $i$th site in a chain, the Ising state written as $\ket{\sigma_1,\ldots,\sigma_{L+1}}$ is disordered.
However, defining spin-1 Ising (or $Z_3$) variables as $m_{i} =  \sigma_{i}-\sigma_{i+1}$,
one finds that $\ket{m_1,\ldots,m_L}$ has the hidden antiferromagnetic order,
i.e.,
$\ket{m_1,\ldots,m_L}$ is identical to the N\'eel ordered state $\ket{\ldots,+1,-1,+1,-1,\ldots}$ if we skip all $m_i=0$.
While two N\'eel states, $\ket{+1,-1,\ldots}$ and $\ket{-1,+1,\ldots}$, are identified by the boundary spin on the first site,
the hidden antiferromagnetic ordered states
are identified by the two boundary $Z_2$ variables, $\sigma_1$ and $\sigma_{L+1}$.

The emergence of the boundary degrees of freedom is
an important topological property.
In fact, 
the exact solution in the Affleck-Kennedy-Lieb-Tasaki(AKLT) model\cite{PRL.59.799},
which is one of rigorous studies of Haldane systems,
shows four-fold degeneracy of the ground states with the Haldane gap in the open boundary condition(OBC),
while there is a unique ground state in the periodic boundary condition(PBC) as Haldane conjectured.
This four($4=2\times 2$)-fold degeneracy is due to the hidden $Z_2\times Z_2$ symmetry.
This kind of topological property, emergence of the edge/surface modes, 
is not only theoretical concept
but also realized experimentally in Haldane systems\cite{PRL.65.3181},
quantum Hall systems\cite{PRL.45.494}, quantum spin Hall systems\cite{SCIENCE.314.1757}, 
and topological insulators\cite{NATURE.452.970}.
Especially,
edge modes in the Haldane system have been applied to the quantum computation\cite{PRL.105.040501}
based on the topological entanglement.
As a theoretical study on the topological entanglement,
AdS/CFT correspondence\cite{PRL.96.181602} is also a hot topic.

Many theoretical models generalized from the AKLT model by means of valence bond solid(VBS) construction\cite{CMP.175.565}
have been studied in higher dimension and/or in various spin symmetries except for $SU(2)$.
However, even in the one-dimensional Heisenberg model with $SU(2)$ symmetry,
what kind of hidden order exists in higher-spin chains is still unclear.
For example, as shown in a rigorous study about the VBS state\cite{JPCM.4.7469},
the hidden $Z_2 \times Z_2$ symmetry breaks down only for odd integer $S$ but remains unbroken for even integer.
As discussed in the recent numerical study on the $S=2$ anisotropic chain\cite{AX.1011.6568},
the determination of the phase diagram
is still worthwhile especially  about existence of the intermediate-$D$(I$D$) phase.
In this sense, higher-$S$ generalization is not straightforward.

\begin{figure}
  \centering
  \resizebox{6cm}{!}{\includegraphics{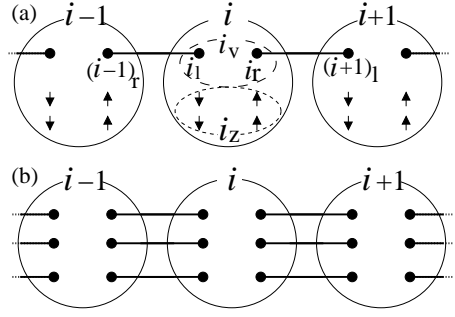}}
  \caption{\label{fig:1} (a)Intermediate-$D$(I$D$) VBS state
    $\mps{A}_{i-1}\mps{A}_{i}\mps{A}_{i+1}$ connected by
    one valence bond and (b) Haldane state connected by $S$ valence
    bonds in the $S=3$ case as two kinds of generalized spin-1 Haldane state.
    Indexes $i_{\rm z},i_{\rm v},i_{\rml},i_{\rm r}$ for decomposed
    spins at $i$th site are defined in the main text.  }
\end{figure}

In this letter, we study what is a natural extension 
toward higher integer spin
based on the Kennedy-Tasaki(KT) transformation $\hat{U}$ \cite{PRB.45.304},
which is nonlocal unitary transformation 
revealing the hidden $Z_2 \times Z_2$ symmetry directly.
In short, the transformed Hamiltonian $\widetilde{H}=\hat{U} \hat{H} \hat{U}^{-1}$
has clear $Z_2 \times Z_2$ symmetry
and the non-local string order in $\hat{H}$ corresponds to conventional ferromagnetic order in $\widetilde{H}$.
In addition,
the four-fold degeneracy which is hidden in $\hat{H}$ with the PBC
corresponds to the four ferromagnetic ground states
in $\widetilde{H}$.
Our motivation is to obtain four ferromagnetic ground states generalized to higher-$S$.
It is surprising that
what this strategy showed us in higher-$S$ models
is not the Haldane state 
but the I$D$ VBS state\cite{JPCM.4.7469}.
(See Fig.~\ref{fig:1}.)
It is quite natural 
if we adopt the concept that the $Z_2$ symmetry is originated from the $Z_2$ variables on the boundaries
and, in this sense,
the Haldane state must have the (hidden) $Z_{S+1} \times Z_{S+1}$ symmetry.

To show the beautiful mathematical structure of our result,
this letter is organized inversely as follows.  First, we start with
the known spin-$S$ VBS Hamiltonian for I$D$ phase\cite{JPCM.4.7469}
as a generalization of the spin-1 AKLT Hamiltonian.
Then,
we write down its ground states as a matrix product state(MPS) with the two boundary variables.
Using a spin decomposition, 
we summarize our results for the KT transformation
and discuss the role of the KT transformation as the topological disentangler,
generalizing the spin-1 topological disentangler studied by Okunishi recently\cite{AX.1011.3277}.
In this meaning,
the KT transformation for arbitrary integer spin
deeply relates not only the hidden $Z_2 \times Z_2$ symmetry and but also the topological entanglement.

\section{Definition:Hamiltonian}
Let us illustrate the I$D$ VBS state for arbitrary spin $S$ in the Schwinger boson picture,
where one spin-$S$ is decomposed into $2S$ spin-1/2's at each site.
To consider the I$D$ VBS states which correspond to the four-fold degenerated ground states,
we limit ourselves to the case that
the number of the valence bonds between the two nearest neighbor sites
is {\it only one}.
As shown in Fig.~\ref{fig:1}(a),
at each site $i$, one spin-1/2 at $i_{\rml}$ couples to the left site,
one spin-1/2 at $i_{\rm r}$ couples to the right site,
and there are $(S-1)$-pairs of up and down spin-1/2's for all the rest in $i_{\rm z}$.
The Hamiltonian which has the I$D$ VBS states as the exact ground states
is given only with the nearest neighbor interactions
as $\hat{H}_{} = \sum_{i} \hat{H}_{i,i+1}$.
Using the spin-$S$ operator at $i$th site, $\hat{\vec{S}}_i$,
and the local $S^z$ basis given by $\hat{S}^z_i\ket{m}_i=m\ket{m}_i$ with  $\hbar=1$,
$\hat{H}_{i,i+1}$ is given as follows:\cite{JPCM.4.7469}
\begin{eqnarray}
\hat{H}_{ij} &=& \hat{P}_{ij}^{(2S)} + \hat{H}^{loc}_{i}+\hat{H}^{loc}_{j}
\label{eq:def:Hij}
,
\\
\hat{P}_{ij}^{(k)} &=& \prod_{m=0}^{k-1} \left[{\left(\hat{\vec{S}}_i + \hat{\vec{S}}_j\right)^2 - m(m+1) \over k(k+1)-m(m+1) }\right]
\label{eq:def:Pij}
,
\\
\hat{H}^{loc}_{i} &=& 1- \sum_{m=-1}^1 \ket{m_i}\bra{m_i}
.
\end{eqnarray}
The symmetrizing projection $\hat{P}_{ij}^{(k)}$ is usually written in $\sum_m \alpha_m (\hat{\vec{S}}_i\cdot \hat{\vec{S}}_j)^m$,
where coefficients $\alpha_m$ are easily obtained due to $(\hat{\vec{S}}_i)^2=S(S+1)$.
Here the notation of local states is defined as $\ket{m_i}=\ket{m}_i$ at $i$th site,
including notations like $\ket{0_i}$ or $\ket{\ua_{i_{\rml}}}$.
Local operators such as $\ket{m_i}\bra{m_i}$ can be extended to operators for the total Hilbert space,
$1\otimes1\otimes\ldots\otimes\ket{m_i}\bra{m_i}\otimes \ldots \otimes 1$,
automatically depending on the context.
The Hamiltonian has  a flexibility: 
one can multiply positive coefficient to $\hat{H}^{loc}_{i}$
and replace it with
\begin{math}
\hat{H}^{loc}_{i}= (S^z_i)^2\left\{ (S^z_i)^2 -1 \right\}
\end{math}
including the negative large-$D$ term, $D \sum_i (S^z_i)^2$.

\section{Definition:MPS}
The I$D$-VBS states are given in the Schwinger boson representation\cite{JPCM.4.7469}.
Explicitly, the states can be written in the MPS form
as
\begin{math}
\mps{\PsiA}\tot = \prod_{i=1}^L \mps{A}_i
\end{math}
with the local matrix
\begin{eqnarray}
\mps{A}_i =  \left(
  \begin{array}{cc}
    -\sqrt{S} \ket{0}_i & \sqrt{S+1} \ket{1}_i \\
    -\sqrt{S+1} \ket{-1}_i& \sqrt{S} \ket{0}_i \\
  \end{array}
\right)/\sqrt{4S-2}  
.
\label{eq:define:A}
\end{eqnarray}
$\mps{\PsiA}\tot$ is a $2\times 2$ matrix corresponding to the 4-fold degenerated ground states,
which are unnormalized and non-orthogonal.
It is easy to check $\hat{H}_{ij} A_iA_{j}$ becomes the zero matrix,
which means that the ground states are zero-energy states.
Since the Hamiltonian is written as sum of the projections,
the other excited states have non-zero eigen-values.
In the $S=1$ case, due to $\hat{H}^{loc}_{i}=0$,
\refeq{eq:def:Hij} becomes
$\hat{H}_{ij}=\hat{P}_{ij}^{(2)}$, which is nothing but the original AKLT Hamiltonian\cite{PRL.59.799}.
In fact, $A_i$ for $S=1$ corresponds to that of the spin-1 Haldane state\cite{JPA.27.6443,JPCM.4.7469}.

Under the OBC, the four-fold degenerated ground states are written as
\begin{eqnarray}
\ket{\Psi_{\sigma_1,\sigma_{L+1}}} = (\mps{\PsiA}\tot)_{\sigma_1,\sigma_{L+1}}
\label{eq:OBCPsi}
,
\end{eqnarray}
with $\left(\sum_{i=1}^{L-1} \hat{H}_{i,i+1}\right) \ket{\Psi_{\sigma_1,\sigma_{L+1}}}=0$.
However, under the PBC, $\hat{H}=\hat{H}_{L,1}+ \sum_{i=1}^{L-1} \hat{H}_{i,i+1}$
requires $\hat{H}_{L,1} \mps{A}_{L} \mps{A}_1=0$.
Then, the unique ground state is written as
\begin{math}
\ket{\Psi_{PBC}} = \Tr \mps{\PsiA}\tot
= \ket{\Psi_{\ua\ua}}+\ket{\Psi_{\da\da}}
.
\end{math}
In general, as \"Ostlund and Rommer introduced\cite{PRL.75.3537},
the ground state is written with the boundary matrix $B$
as,
\begin{math}
\ket{\Psi(B)} =   \Tr \mps{\PsiA}\tot B
=\sum_{\sigma_1,\sigma_{L+1}}
(B)_{\sigma_{L+1},\sigma_1} \ket{\Psi_{\sigma_1,\sigma_{L+1}}}
,
\end{math}
and $\ket{\Psi(1)}=\ket{\Psi_{PBC}}$ for the identity matrix.
Even in the PBC,
importance of this boundary matrix has been noticed again recently in the numerical\cite{AX.1011.0576} and rigorous studies\cite{JPA.43.175003,JPSJ.79.073002}.

\section{Result:four classical state}
Higher-spin generalization of the KT transformation is given as\cite{JPCM.4.7469}
\begin{eqnarray}
\UKT\tot = \prod_{i<j}\e^{\i\pi \hat{S}^z_i \hat{S}^x_j} = {\UKT\tot}^\dagger = (\UKT\tot)^{-1}
,
\end{eqnarray}
with the spin-$S$ operator $\hat{S}^\alpha_i$.
The hidden $Z_2\times Z_2$ symmetry is revealed by $\UKT\tot$
because the ferro magnetic correlation function $\hat{S}^\alpha_i
\hat{S}^\alpha_j$
is transformed into
the string operator for $\alpha=x,z$\cite{JPCM.4.7469}.
For example, 
\begin{math}
\UKT\tot
\hat{S}^z_i
\hat{S}^z_j
(\UKT\tot)^{-1}
=
\hat{S}^z_i
\e^{\i\pi \sum_{k=i}^{j-1} \hat{S}^z_k}
\hat{S}^z_j
\end{math}
in arbitrary integer $S$.

As the main result of this letter,
we will write down the four states after the KT transformation.
It is interesting that we can obtain a simple form if we use the boundary matrix $B=\Omega$
defined as
\begin{eqnarray}
\Omega =\left(
  \begin{array}{cc}
    1&1\\
    -1&1
  \end{array}\right)
.
\end{eqnarray}
Then, the MPS after the KT transformation can be written as
\begin{eqnarray}
\UKT\tot \mps{\PsiA}\tot \Omega = \left(
  \begin{array}{cc}
    (-1)^L & (-1)^L\e^{\i \pi \hat{S}^z\tot}\\
    -\e^{\i \pi \hat{S}^x\tot} & \e^{\i \pi \hat{S}^z\tot}\e^{i\pi\hat{S}^x\tot}\\
  \end{array}
\right)\ket{\phi}
\label{eq:UKTPsi}
,
\end{eqnarray}
with $\hat{S}^\alpha\tot=\sum_{i=1}^L \hat{S}^\alpha_i$
and
\begin{eqnarray}
\ket{\phi} =
\prod_i\left( {\sqrt{S}\ket{0}_i+(-1)^{(S+1)(i+L)}\sqrt{S+1}\ket{1}_i\over \sqrt{4S-2} }\right)  
\label{eq:def:phi}
.
\end{eqnarray}
These four Ising states are generalization of ferromagnetic states in the $S=1$ case\cite{PRB.45.304}.
Using this formula,
one can easily transform an MPS with any boundary.
For example, in the PBC for even $L$, one can obtain
\begin{math}
 \UKT\tot \ket{\Psi_{PBC}} = 
 \Tr[ (\UKT\tot \mps{\PsiA}\tot \Omega) \Omega^{-1} ]= 
(1+ \e^{\i \pi \hat{S}^z\tot})(1+ \e^{\i \pi \hat{S}^x\tot})
\ket{\phi}/2
.
\end{math}

The $2\times 2$ matrix in \refeq{eq:UKTPsi}
reveals the hidden $Z_2 \times Z_2$ symmetric operations about 180$^\circ$ rotation via $x$ and $z$-axis
for four (anti)ferromagnetic states
thanks to the boundary matrix $\Omega$.
Note that these four  states are written in the classical (or Ising) states,
i.e.,
the direct-product states of the local states.
In this meaning, $\UKT\tot$ is the total disentangler.
That is, the KT transformation can disentangle each valence bond singlet having the $\log 2$ entanglement entropy
corresponding to the $Z_2$ variable,
$\sigma_i$.
In the following,
we will illustrate it  generalizing Okunishi's result \cite{AX.1011.3277}
to higher-$S$ in a unified way using a spin decomposition.

\section{Definition:spin decomposition}
Let us illustrate the spin decomposition.
The spin-$S$ operator at $i$th site can be decomposed into spin-$(S-1)$ operator $\hat{\vec{S}}_{i_{\rm z}}$ 
and spin-$1$ operator $\hat{\vec{S}}_{i_{\rm v}}$
when we introduce the relation between the local basis
as
\begin{eqnarray}
\ket{m}_i =\sum_{m'm''}\ket{m'}_{i_{\rm v}} \ket{m''}_{i_{\rm z}} \braket{1,m';S-1,m''}{S,m}  
\label{eq:CG},
\end{eqnarray}
where $\braket{J,m}{J',m';J'',m''}$ are Clebsh-Gordan coefficients.
With using the projection operator defined in \refeq{eq:def:Pij},
the equation for spin operators is written as
$\hat{\vec{S}}_i=\hat{P}_{i_{\rm z}i_{\rm v}}^{(S)} ( \hat{\vec{S}}_{i_{\rm z}}+\hat{\vec{S}}_{i_{\rm v}})$.
Here we note our ambiguity in this equation: the r.h.s. 
must be $2S+1$ dimensional matrices with the Hilbert space spanned by $\ket{-S}_i,\ldots,\ket{S}_i$,
but, the l.h.s. equals to the product of $3(2S-1)$ dimensional matrices with direct-product states $\ket{m}_{i_{\rm v}}\ket{m'}_{i_{\rm z}}$ 
for $|m|\leq 1$ and $|m'| \leq S-1$.
In this case, $\ket{m}_i$ must be written in $\ket{m}_{i_{\rm v}}\ket{m'}_{i_{\rm z}}$ via \refeq{eq:CG}.
In the following, the Hilbert space of each equation is not explicitly defined 
and automatically changed depending on the context.
The same ambiguity in the identical equation $\hat{\vec{S}}_i=\hat{P}_{i_{\rm z}i_{\rm v}}^{(S)} ( \hat{\vec{S}}_{i_{\rm z}}+\hat{\vec{S}}_{i_{\rm v}})
(\hat{P}_{i_{\rm z}i_{\rm v}}^{(S)})^\dagger$
might be removed by rewriting the projection operator defined in \refeq{eq:def:Pij} as
\begin{math}
\hat{P}_{i_{\rm z}i_{\rm v}}^{(S)}=\sum_{mm'm''} \ket{m_{i}} \braket{S,m_{i}}{1,m'_{i_{\rm v}};S-1,m''_{i_{\rm z}}} \bra{m'_{i_{\rm v}}m''_{i_{\rm z}}}
.
\end{math}
Another useful formula is 
\begin{math}
\hat{\vec{S}}_i \hat{P}_{i_{\rm z}i_{\rm v}}^{(S)} =\hat{P}_{i_{\rm z}i_{\rm v}}^{(S)} \left( \hat{\vec{S}}_{i_{\rm z}}+\hat{\vec{S}}_{i_{\rm v}}\right)
.
\end{math}
This is the spin decomposition in the opposite of the spin composition $\hat{\vec{S}}_{ij}=\hat{\vec{S}}_i + \hat{\vec{S}}_j$.
For readability, we summarize notations of operators.
All operators have the hat notation, $\hat{\;\;\;}$, and subscripts denoting the region where the operator is defined:
single site operator $\hat{S}^z_i$ and two site operator $\hat{H}_{ij}$,
except for total operator $\hat{S}^z\tot$.
This rule for subscript is also applied to the MPS, such as $\mps{A}_i,$ and $ \mps{\PsiA}\tot$.

Using the spin decomposition,
the local MPS $\mps{A}_i$ for $S>1$ can be constructed from spin-1 MPS $\mps{A}_{i_{\rm v}}$ as
\begin{eqnarray}
\mps{A}_i &=& \hat{P}_{i_{\rm z}i_{\rm v}}^{(S)} \ket{0}_{i_{\rm z}}\mps{A}_{i_{\rm v}}
\label{eq:decom:i2zv}
.
\end{eqnarray}
In addition, we can decompose  a spin-$S$ into spin-1/2's.
This is nothing but the schematic picture of I$D$-VBS state\cite{CMP.175.565} as shown in Fig.~\ref{fig:1}(a).
For simplicity, we decompose a spin-1 only at $i_{\rm v}$-site into two spin-1/2's
as
\begin{eqnarray*}
\mps{A}_{i_{\rm v}} =   \hat{P}_{i_{\rml}i_{\rm r}}^{(1)} 
\left(
  \begin{array}{c}
    \ket{\ua}_{i_{\rml}} \\
    \ket{\da}_{i_{\rml}} \\
  \end{array}
\right)
\left(\ket{\ua}_{i_{\rm r}},\ket{\da}_{i_{\rm r}}\right)
\e^{\i \pi \sy}
.
\end{eqnarray*}
Here we use the $2\times 2$ matrix $\sop^y = {1\over 2}\left(
  \begin{array}{cc}
    0&\i\\
    -\i&0
  \end{array}\right)
$, which is the matrix representation of the spin-1/2 operator.
A product of the matrices, $\mps{A}_{i_{\rm v}}\mps{A}_{j_{\rm v}}$
includes the valence bond singlet
\begin{eqnarray*}
\ket{\singlet}_{{i_{\rm r}},{j_{\rml}}}
=
\left(\ket{\ua}_{i_{\rm r}},\ket{\da}_{i_{\rm r}}\right)
\e^{\i \pi \sy}
\left(
  \begin{array}{c}
    \ket{\ua}_{j_{\rml}} \\
    \ket{\da}_{j_{\rml}} \\
  \end{array}
\right)/\sqrt{2}.
\end{eqnarray*}

\section{Discussion:boundary variables}
Before we mention the disentangler,
we summarize the $Z_2$ property of the spin decomposed MPS,
recalling the introduction of this letter.
The hidden antiferromagnetic order and $Z_2\times Z_2$ symmetry 
come from the random variables, $\sigma_i$. 
These spin-1/2 Ising (or $Z_2$) variables correspond to the artificial  degrees of the freedom
of the matrix space as
\begin{math}
\ket{\Psi_{\sigma_1,\sigma_{L+1}}} = \sum_{\sigma} (\mps{A}_1)_{\sigma_1\sigma_2}  
(\mps{A}_2)_{\sigma_2\sigma_3}  
\cdots
(\mps{A}_L)_{\sigma_L\sigma_{L+1}}
,
\end{math}
which is defined in \refeq{eq:OBCPsi}.
After the spin decomposition in \refeq{eq:decom:i2zv},
a matrix element $(\mps{A}_{i_{\rm v}})_{\sigma_i\sigma_{i+1}}$
is proportional to
\begin{math}
\ket{\sigma_i}_{i_{\rml}}  \ket{-\sigma_{i+1}}_{i_{\rm r}}  
\end{math}
in the spin decomposed basis.
In the original spin basis,
$(\mps{A}_{i})_{\sigma_i\sigma_{i+1}}$ corresponds to
\begin{math}
\ket{m_i}_i
\end{math}
with 
\begin{math}
m_i =\sigma_i-\sigma_{i+1}
.
\end{math}
The correspondence can be checked from \refeq{eq:define:A}.
This is the explicit correspondence to the RSOS model\cite{PRB.40.4709}
as already mentioned in this letter.
We emphasize that
$\sigma_i$ is the source of the $\log 2$ entanglement entropy
of the valence bond singlet 
$\ket{\singlet}_{{i_{\rm r}},{j_{\rml}}}$,
which is the target of the disentangler.

\section{topological disentanglement(TD)}
\begin{figure}
  \centering
  \resizebox{9cm}{!}{\includegraphics{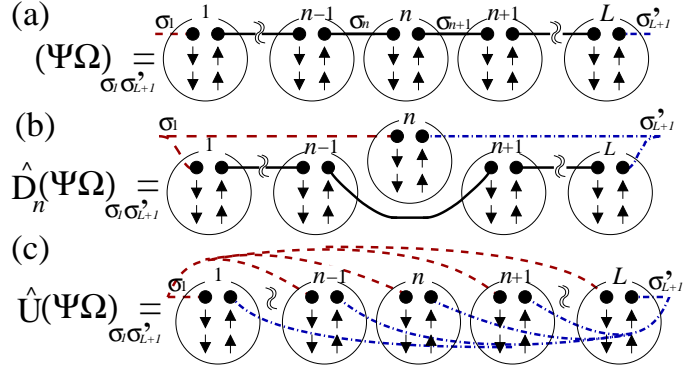}}
  \caption{\label{fig:2}(color online) Schematic diagram for (a) Intermediate-$D$ VBS state $\mps{\PsiA}\tot\Omega$,
    (b) partially disentangled state via $\hat{D}_n$,
    and (c) totally disentangled state after the KT transformation $\hat{U}$.
  }
\end{figure}
The topological disentangler\cite{AX.1011.3277}, defined as 
\begin{eqnarray}
\hat{D}_n = \prod_{i=1}^{n-1} \e^{\i\pi \hat{S}^z_i \hat{S}^x_n}
\prod_{i=n+1}^{L}\e^{\i\pi \hat{S}^z_n \hat{S}^x_i} 
,
\end{eqnarray}
has the same form in the higher-spin generalization.
Then, using the boundary $Z_2$ variables $\mu=\sigma_1,\nu=\sigma'_{L+1}$
and the total projection \begin{math}
\hat{P}\tot =  \prod_{i=1}^L \hat{P}_{i_{\rm z}i_{\rm v}}^{(S)} \hat{P}_{i_{\rml}i_{\rm r}}^{(1)}
\end{math}
, we have
\begin{eqnarray*}
(\mps{\PsiA}\tot  \Omega)_{\mu\nu}
=
- \sqrt{2^L}
\hat{P}\tot
\ket{\mu}_{1_{\rml}}
\prod_{i=1}^{L-1}\ket{\singlet}_{{i_{\rm r}},{(i+1)_{\rml}}}
\ket{\nu}_{L_{\rm r}}^x
\prod_{i} \ket{0_{i_{\rm z}}}
,
\end{eqnarray*}
where 
\begin{math}
\ket{\sigma}^x_{k} =\e^{-\i {\pi\over 2} \hat{S}^y_k} \ket{\sigma}_{k}
\end{math}
for $k=i_{\rml}$ or $k=i_{\rm r}$
are local spin-1/2 states in spin-$x$ basis.
Since we can write $(\mps{\PsiA}\tot  \Omega)_{\sigma_1\sigma'_{L+1}}
=\sum_{\sigma_{L+1}} (\mps{\PsiA}\tot)_{\sigma_1\sigma_{L+1}} \times (\Omega)_{\sigma_{L+1}\sigma'_{L+1}}
=\sum_{\sigma_{L+1}} \ket{\Psi_{\sigma_1\sigma_{L+1}}}  (\Omega)_{\sigma_{L+1}\sigma'_{L+1}}$,
$\Omega$ is a transformation (or mapping) between $\sigma_{L+1}$ and $\sigma'_{L+1}$.
In fact, due to $\Omega=\sqrt{2}\e^{\i {\pi\over 2} \sy}$,  $\sigma'_{L+1}$ is a variable in
spin-$x$ axis.  This is the meaning of $\Omega$ and the local basis at $L_{\rm r}$ is written in $\ket{\nu}^x_{L_{\rm r}}$.
The local states on both boundaries are determined by $\mu=\sigma_1,\nu=\sigma'_{L+1}$,
but each singlet is entangled and has a free (or random) variable $\sigma_i$ as shown in Fig.~\ref{fig:2}(a).
For odd $S$, this MPS is disentangled as
\begin{eqnarray}
&&\nonumber
\hat{D}_n (\mps{\PsiA}\tot  \Omega)_{\mu\nu} =
\sqrt{2^L}
(\Omega)_{\mu\nu}
\hat{P}\tot
\ket{\mu}_{1_{\rml}}\ket{\mu}_{n_{\rml}}\ket{\nu}^x_{n_{\rm r}}\ket{\nu}^x_{L_{\rm r}}
\\&&
\times
\ket{\singlet}_{(n-1)_{\rm r};(n+1)_{\rml}}
\prod_{i\neq n-1,n} \ket{\singlet}_{i_{\rm r};(i+1)_{\rml}}
\prod_{i} \ket{0}_{i_{\rm z}}
,
\label{eq:disentangle:D}
\end{eqnarray}
and
\begin{eqnarray}
\hat{U} (\mps{\PsiA}\tot  \Omega)_{\mu\nu}  
=
(-\sqrt{2})^L
(\Omega)_{\mu\nu}^{L+1}
\hat{P}\tot
\prod_i \ket{\mu}_{i_{\rml}}\ket{\nu}^x_{i_{\rm r}} \ket{0}_{i_{\rm z}}
\label{eq:disentangle:U}
,
\end{eqnarray}
as shown in Fig.~\ref{fig:2}.
In short,
the target local states are determined by the boundary variables $\mu,\nu$.

Equations~(\ref{eq:disentangle:D}) and (\ref{eq:disentangle:U}) are valid only for odd $S$.
For general $S$, we can obtain the same result if we replace $\e^{\i\pi\hat{S}^z_i  \hat{S}^x_j}$
with $\e^{\i\pi\hat{S}^z_i (\hat{S}^x_j+S-1)}$
in the definition of $\hat{U}$ and $\hat{D}_n$.
This is the reason why $\ket{\phi}$ in \refeq{eq:def:phi}
has antiferromagnetic-like alternating behavior for even $S$.
These results can be proved easily by using the spin decomposition\cite{comment:1},

\section{Discussion:Haldane case}
It is instructive to show how the KT transformation fails to disentangle the Haldane state for $S>1$.
For example, let us consider the Haldane state for $L=2,$ and $S=2$.
We can decompose a $S=2$ spin at $i$th site into two spin-1's at $i_{\rm r}$
and $i_{\rml}$
with
the valence bond singlet 
\begin{math}
\ket{\singlet}_{i_{\rm r},(i+1)_{\rml}}
=
{\ket{1}_{i_{\rm r}}\ket{-1}_{(i+1)_{\rml}}
-\ket{0_{i_{\rm r}}}\ket{0}_{(i+1)_{\rml}}
+\ket{-1}_{i_{\rm r}}\ket{1}_{(i+1)_{\rml}}
\over \sqrt{3}}
,
\end{math}
which is the unique ground state of the Hamiltonian $\vec{S}_{i_{\rm r}}\cdot \vec{S}_{(i+1)_{\rml}}$
and has $\log 3$ entanglement entropy.
Then, one of $3\times 3$ states, $\ket{\Psi_{00}}
=
\hat{P}_{\rm tot}\ket{0}_{1_{\rml}}   
\ket{\singlet}_{1_{\rm r},2_{\rml}}
\ket{0}_{2_{\rm r}}   
$,
is transformed as
\begin{math}
\e^{\i\pi \hat{S}^z_1 \hat{S}^x_2}
\ket{\Psi_{00}}
=
\hat{P}_{\rm tot}\ket{0}_{1_{\rml}}
{\ket{1}_{1_{\rm r}}\ket{1}_{2_{\rml}}
-\ket{0}_{1_{\rm r}}\ket{0}_{2_{\rml}}
+\ket{-1}_{1_{\rm r}}\ket{-1}_{2_{\rml}}
\over \sqrt{3}}
\ket{0}_{2_{\rm r}}
.
\end{math}
As shown in this example,
the KT transformation
fails to disentangle the singlet corresponds to $Z_3$ symmetry.

\section{conclusion}
In summary, we have rigorously found that 
four-fold degenerated Ising-like states generalized to arbitrary integer spins
correspond to four-fold degenerated I$D$-states
via the KT transformation as the total disentangler.
In the view point of the one-site disentangler,
we have given the higher-spin generalization of Okunishi's paper\cite{AX.1011.3277}
using the spin decomposition representation
as an alternative to the Schwinger boson representation.

The spin decomposition approach reminds us the decomposition
of a $S=1$ spin into ferro-magnetically coupled $S=1/2$ spins, 
discussed in the $S=1/2$ quantum spin chain with bond alternation\cite{PRB.45.2207,PRB.46.3486}.
The ground state is adiabatically connected to the direct product of the local singlets.
Adding $\ket{0}_{i_{\rm z}}$ to it, we can make a corresponding model
with local singlets.
However, 
such local singlets can be disentangled easily by a unitary operator, such as $\e^{\i\pi\hat{S}^z_{i_{\rml}} \hat{S}^x_{i_{\rm r}}}$.
This means the entanglement of the singlet can easily destroyed by a two-site unitary operator,
while we need the nonlocal KT transformation to disentangle the singlet in the spin-$S$ chain.
This is because we have operators like $\hat{S}^\alpha_i = \hat{P} (\hat{S}^\alpha_{i_{\rm r}}+\hat{S}^\alpha_{i_{\rml}}+\hat{S}^\alpha_{i_{\rm z}})\hat{P}^\dagger$ only.
In this sense, a highly non-trivial task is to obtain the disentangler,
and it is surprising that the disentangler for the VBS states in the $S=1$ ALKT model is the KT transformation\cite{AX.1011.3277}.
Moreover, as written in this letter,
the higher-spin generalization can be obtained
and leads us to the intermediate-$D$ states with $Z_2$ boundary variables.
Since the KT transformation fails to disentangle the Haldane state for $S>1$,
we think this way is a natural extension in 
the sense that the KT transformation can play a role of the disentangler of the $\log 2$ entanglement entropy.

\section{$Z_2$ Berry phase}
This Hamiltonian breaks the spin $SU(2)$ symmetry but has $U(1)$
rotational symmetry via spin $z$-axis.
Then, we adopt the usual $Z_2$ Berry phase via gauge twist on $z$-axis\cite{PRB.77.094431},
which corresponds to the twisted boundary condition
used in the level spectroscopy in the $S=2$ chain\cite{AX.1011.6568}.
Since the gauge twist just modifies the boundary matrix,
we can prove that the I$D$ state gives non-trivial $\pi$ Berry phase due one singlet on the bond.

\section{$q$ deformation}
As a future problem, one can consider the $q$-deformed model with $U_q(su(2))$ quantum group\cite{JPA.27.6443}.
In addition, construction of the disentangler corresponding to the hidden $Z_{S+1}\times Z_{S+1}$ symmetry 
is still an open question, but this (dis)entanglement view of point will be important in this generalization.

\acknowledgement\;\;\;
We are grateful to K. Okunishi for showing us his results\cite{AX.1011.3277} prior to publication.
This work was supported in part by a Grant-in-Aid (No. 20740214), and Global COE Program (Core Research and Engineering of Advanced Materials-Interdisciplinary Education Center for Materials Science) from the Ministry of Education, Culture, Sports, Science and Technology of Japan.

\bibliography{macro,wiki,book,comment}

\begin{thebibliography}{10}
\bibitem{comment:1}
For example, one can use \begin{math}\e^{\i\pi\hat{\Stmp}^z_1
  \hat{\Stmp}^x_2}\hat{Q} \ket{z} =\hat{Q} \ket{z}
  \e^{\i\pi\left(\hat{\Stmp}^z_{1_{\rm z}}+ \hat{\Stmp}^z_{1_{\rm v}}\right)
  \left(\hat{\Stmp}^x_{2_{\rm z}}+ \hat{\Stmp}^x_{2_{\rm v}}\right)} = \hat{Q}
  \ket{z} \e^{\i\pi \hat{\Stmp}^z_{1_{\rm v}} (\hat{\Stmp}^x_{2_{\rm
  v}}+\Stmp-1)} , \end{math} and $ \e^{\i\pi\hat{\Stmp}^z_1
  (\hat{\Stmp}^x_2+\Stmp-1)} \hat{Q}\ket{z} = \hat{Q} \e^{\i\pi
  \hat{\Stmp}^z_{1_{\rm v}} \hat{\Stmp}^x_{2_{\rm v}}}\ket{z} , $ where
  $\hat{Q}=\hat{P}^{(\Stmp)}_{{1_{\rm z}}{1_{\rm v}}}\hat{P}^{(\Stmp)}_{{2_{\rm
  z}}{2_{\rm v}}}$ and $\ket{z}=\ket{0_{1_{\rm z}}0_{2_{\rm z}}}$.
\end{thebibliography}

\newpage
\end{document}